\documentclass[10pt,journal]{IEEEtran}

\IEEEoverridecommandlockouts

\usepackage{tikz}
\usepackage{textcomp}
\usepackage{lipsum}

\newcommand\copyrighttext{%
  \footnotesize \textcopyright 2019 IEEE. Personal use of this material is permitted.
  Permission from IEEE must be obtained for all other uses, in any current or future
  media, including reprinting/republishing this material for advertising or promotional
  purposes, creating new collective works, for resale or redistribution to servers or
  lists, or reuse of any copyrighted component of this work in other works.
  DOI:{10.1109/TASC.2019.2892584}}
\newcommand\copyrightnotice{%
\begin{tikzpicture}[remember picture,overlay]
\node[anchor=north,yshift=-10pt] at (current page.north) {\fbox{\parbox{\dimexpr\textwidth-\fboxsep-\fboxrule\relax}{\copyrighttext}}};
\end{tikzpicture}%
}

\usepackage{cite}

\ifCLASSINFOpdf

\else
  \fi

\usepackage{amsmath,amssymb}

\usepackage{array}
\usepackage{xspace}
\usepackage{color}
\usepackage{etoolbox}

\newcommand{\NbSn}{Nb\textsubscript{3}Sn}
\newcommand{\Tc}{$T_c$}
\newcommand{\Zs}{$Z_s$}
\newcommand{\Q}{$Q$}
\newcommand{\f}{$\nu_{res}$}
\newcommand{\TEmode}{TE\textsubscript{011}}
\newcommand{\Celsius}{$^\circ$C}

\begin{document}
\bstctlcite{IEEEexample:BSTcontrol}

\title{Surface Impedance Measurements on \NbSn~in High Magnetic Fields}
\author{Andrea~Alimenti,~\IEEEmembership{Student Member,~IEEE,}
        Nicola~Pompeo,~\IEEEmembership{Senior Member,~IEEE,}
        Kostiantyn~Torokhtii,~\IEEEmembership{Member,~IEEE}
        Tiziana~Spina, Ren\'{e}~Fl\"{u}kiger,
        Luigi~Muzzi,~\IEEEmembership{Senior Member,~IEEE}
        Enrico~Silva,~\IEEEmembership{Senior Member,~IEEE}
\thanks{A. Alimenti, N. Pompeo, K. Torokhtii and E. Silva are with the Department
of Engineering, Universit\`{a} Roma Tre, 00146 Roma,
Italy. Corresponding author: A. Alimenti; e-mail: andrea.alimenti@uniroma3.it.}%
\thanks{T. Spina and R. Fl\"{u}kiger are with the European Organization for Nuclear Research (CERN), Technology Department, Geneve, Switzerland}%
\thanks{L. Muzzi is  with ENEA - Frascati Research Centre, Frascati, Italy}%
\thanks{Manuscript received January 9, 2019.}}

{}

\maketitle
\copyrightnotice

\begin{abstract}

\NbSn~is a superconductor of great relevance for perspective RF applications. We present for the first time surface impedance ($Z_s$) measurements at 15~GHz and low RF field amplitude on \NbSn~in high magnetic fields up to 12~T, with the aim of increasing the knowledge of \NbSn~behavior in such conditions.  $Z_s$ is a fundamental material parameter that directly gives useful information about the dissipative and reactive phenomena when the superconductor is subjected to high-frequency excitations. Therefore, we present an analysis of the measured $Z_s$ with the aim of extracting interesting data about pinning in \NbSn~ at high frequencies. From \Zs~we extract the vortex motion complex resistivity to obtain the $r$-parameter and the depinning frequency $\nu_p$ 
in high magnetic fields.  The comparison of the results with the literature shows that the measured $\nu_p$ on bulk \NbSn~is several times greater than that of pure Nb. This demonstrates how \NbSn~can be a good candidate for RF technological applications, also in high magnetic fields.
\end{abstract}


\IEEEpeerreviewmaketitle

\section{Introduction}

\IEEEPARstart{C}{urrently} \NbSn~is the most interesting technological superconductor both for high performance dc applications like magnets for nuclear fusion or particle accelerators, and for potential radiofrequency applications as resonating cavities \cite{xu2017review, thome1994design, posen2017nb3sn, valente2016superconducting,padamsee1998rf,flukiger2012overview}.

In applications like superconductive power cables in high magnetic fields, it is well known that vortex motion is the main contribution to the conduction losses. Hence, the experimental study of fluxons behavior, and their pinning, is of great relevance.

At radiofrequency (RF) and microwaves (mw), pinning is still a relevant topic, since vortices are much more free to dissipate. Challenging applications of \NbSn~are represented by the RF accelerating cavities for particles accelerators \cite{posen2017nb3sn,padamsee2009rf}. Nb cavities are currently being used but their technological limits seem to be reached \cite{posen2017nb3sn}, so to overcome their performances it is necessary to consider other superconductors and \NbSn~is a potential candidate \cite{becker2015analysis}. \NbSn~offers approximately twice the critical temperature \Tc~and the superheating field $H_{sh}$ with respect to Nb. This yields an improved cryogenic efficiency and perspective higher accelerating fields \cite{posen2017nb3sn}. However, it must be mentioned that to date, \NbSn\ cavities show a limit of peak surface magnetic field at ${\sim70}$~mT \cite{posen2017nb3sn} which is still lower than the highest peak RF magnetic field reached with bulk Nb cavities ${\sim209}$~mT \cite{grassellino2018accelerating}.

The improvement of the actual technological limits of this superconductor, in terms of critical current $J_c$ and superheating field $H_{sh}$, is a mandatory requirement for the development of some of the presented applications. For this reason an in depth study of the RF electrodynamic response of \NbSn~when subjected to extreme working conditions is needed. 

For RF applications the depinning frequency $\nu_p$ is the most relevant parameter because it 
marks the boundary between the frequency band where the response of fluxons to harmonic excitation is mainly elastic ($\nu<\nu_p$) and the range where the vortex oscillation becomes purely dissipative ($\nu>\nu_p$) \cite{gittleman1966radio}. The higher the $\nu_p$  the higher the usable working frequencies are  with a given SC with reduced dissipation.

Despite the relevance of Nb\textsubscript{3}Sn, to our knowledge no high magnetic fields microwave measurements are present in literature on \NbSn. In this work we present the first microwave characterization of \NbSn~in high magnetic fields through \Zs~measurements, performed with low RF field amplitude. From the \Zs~analysis, the $\nu_p$ is determined as a function of temperature $T$. We find that $\nu_p$ attains values much larger than in Nb, thus making \NbSn~an attractive material for its RF and mw potential performance in high magnetic fields.

The paper is organized as follows. In Sec.\ref{Zs} we briefly recall the main model for $Z_s$ in a superconductor in the vortex state. In Sec.\ref{exp} we describe the experimental setup and method. In Sec.\ref{results} we present the characterization of the sample and the experimental results for $Z_s$. Short conclusions are presented in Sec.\ref{conc}.

\section{Surface impedance in the mixed state}
\label{Zs}

At high frequencies, the electromagnetic response of a conductor is modelled by the complex surface impedance \Zs. For bulk good conductors in the local limit and normal incidence electromagnetic waves, the surface impedance \cite{collin2007foundations} is defined as ${Z_s=R_s+iX_s=\sqrt{i\omega\mu_0\tilde\rho}=i\omega\mu_0\tilde{\lambda}}$, where $R_s$ and $X_s$ are the surface resistance and reactance respectively, $\omega$ the angular frequency, $\mu_0$ the vacuum magnetic permeability and ${\tilde\rho=i\omega\mu_0\tilde\lambda^2}$ the complex resistivity and $\tilde{\lambda}$ a complex shielding length. The magnetic field $H$ and temperature $T$ dependence of $\tilde{\lambda}$ models the dissipative and reactive phenomena of type--II superconductors: quasiparticle scattering and vortex flow, as well as the pinning properties. As defined in \cite{coffey1991unified}, $\tilde{\lambda}$ is a function of the complex conductivity  $\sigma_{2f}=\sigma_1-i\sigma_2$ and of the vortex motion resistivity $\rho_{vm}$ \cite{tinkham1996introduction}. If ${\sigma_1\ll\sigma_2}$, hence not too close to \Tc~and $H_{c2}$, one finds:
\begin{equation}\label{eqn:Zsrho}
Z_s=i\omega\mu_0\sqrt{\lambda^2-i\rho_{vm}/\omega\mu_0},
\end{equation} 
where $\lambda$ is the London penetration depth and $\rho_{vm}$ is the vortex motion complex resistivity.
If the applied magnetic field ${H=0}$ and ${T\rightarrow0}$~K, there are no fluxons in the SC, hence ${\rho_{vm}=0}$ and ${R_{s,ref}\sim0,\;X_{s,ref}=\omega\mu_0\lambda}$.

High--frequency microwave measurements are particularly interesting since at these frequencies the displacements of the fluxons from their equilibrium positions, due to the induced microwave (mw) currents ($J_{mw}$), are so small that dynamic mutual interactions of fluxons can be discarded or reduced to an average effect. Thus, $\rho_{vm}$ can be described by simplified, single--fluxon, local models such as the Gittleman-Rosemblum (GR) model where one writes (having neglected thermal fluctuations)\cite{gittleman1966radio,pompeo2008reliable}:
\begin{equation}
\label{eq:GR}
\rho_{vm}=\rho'_{vm}+i\rho''_{vm}=\rho_{ff}\frac{1}{1-i\frac{\nu_p}{\nu}},
\end{equation}
with $\rho_{ff}$ the flux-flow resistivity, and $\nu_p$ appears explicitly.

Within the GR model the so called $r$-parameter, defined as  ${r=\rho''_{vm}/\rho'_{vm}}$, gives immediately ${r =\nu_p/\nu}$, and thus $\nu_p$ is directly obtained.

\section{Experimental technique}
\label{exp}

Dielectric loaded resonators offer high sensitivity for \Zs~measurement \cite{chen2004microwave}. In this technique the dielectric crystal, loaded into the cavity, is used to focus  the electromagnetic (e.m.) field near the axis of the resonator, thus limiting the conduction losses. The higher the electrical permittivity of the crystal, the more the physical dimension of the resonator are reduced at the same working frequency, useful to probe small samples.

The sample is loaded into the cavity in order to substitute a base of the resonator (end-wall replacement configuration) \cite{chen2004microwave}. We measure the changes in the quality factor \Q~and in the resonating frequency \f~with $T$ or $H$ 
due to the change in surface impedance of the superconducting sample. \Q~and \f~are obtained fitting the complex scattering parameters  ${S_{12}(\nu)}$ or ${S_{21}(\nu)}$. ${S_{11}(\nu)}$ and ${S_{22}(\nu)}$ are used to evaluate that the coupling factors of the 2-ports  ${\beta_1+\beta_2=\beta<0.01}$ \cite{chen2004microwave}, firmly in the undercoupled regime, thus the measured quality factor ${Q_L=(1+\beta)Q\simeq Q}$. 

\Q~and \f~yield  $R_s$ and $X_s$ respectively, by means of the relation: 
\begin{equation}\label{eqn:RsDXs}
R_s+i\Delta X_s=\frac{G_s}{Q}-i2G_s \frac{\Delta \nu_{res}}{\nu_{res}} -background,
\end{equation}
where $G_s$ is a calculated geometrical factor, $\Delta$ indicates a variation with respect to a reference value, and $background$ indicates the (complex and $T$--dependent) contribution given by the resonator itself. A calibration of the resonator with a metallic sample
 allows to remove the background. Once the background is subtracted, the absolute values of  ${Z_s(T,H)}$ are obtained making use of the following fixed points: we set ${R_s(H=0,T\rightarrow0)\sim0}$ (about this choice, see below for sensitivity comments) and  ${R_s=X_s}$ above $T_c$ (real quasiparticle conductivity, Hagen-Rubens limit).

\begin{figure}[!t]
\centering
\includegraphics[width=1\columnwidth]{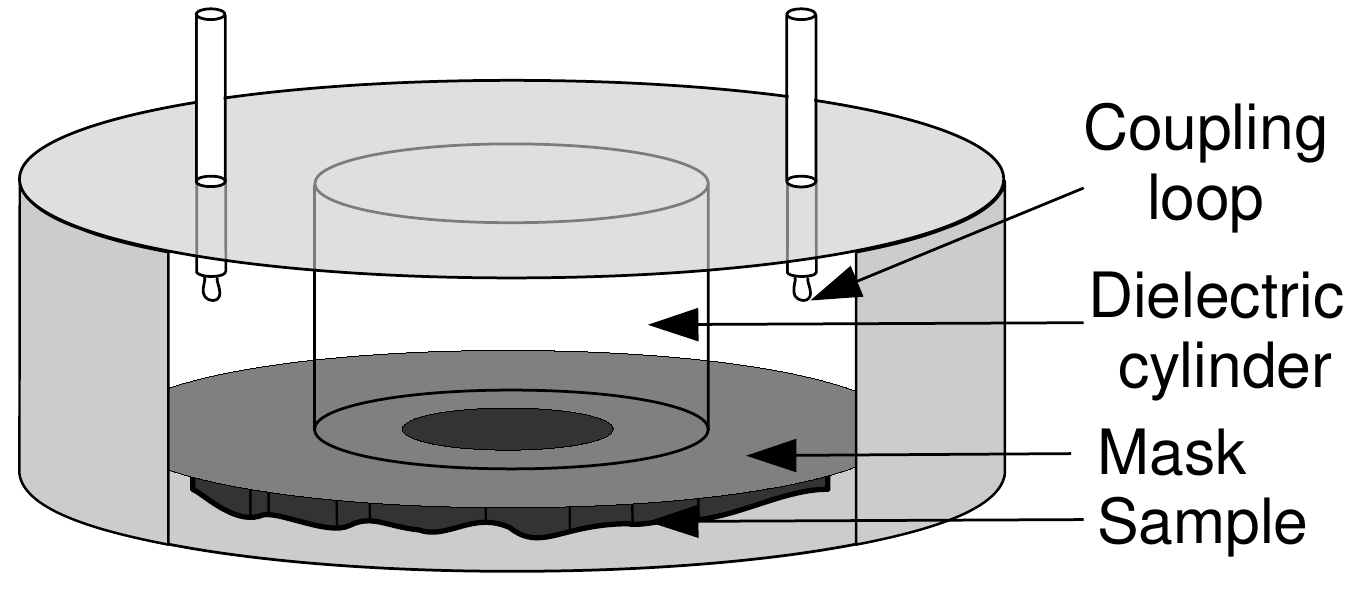}
\caption{Dielectric loaded resonator sketch. The metallic mask is used to keep the cylindrical symmetry despite the irregular shape of the sample. The resonating mode is excited with coaxial cables ended with magnetic loops. The dc magnetic field is perpendicular to the sample flat surface.}
\label{fig:DRsketch}
\end{figure}

The specific dielectric resonator used here is of Hakki-Coleman type \cite{hakki1960dielectric}. The sketch of the resonator is shown in \figurename~\ref{fig:DRsketch}. The entire assembly makes use of several springs in order to avoid issues related to the thermal expansions of the different components. The resonator works in transmission, and it is excited in the \TEmode~mode at ${\sim14.9}$~GHz with coaxial cables terminated with magnetic loops. A single--crystal sapphire cylindrical puck, 5.0~mm height and 8.0~mm diameter, loads the OFHC copper.  Low dielectric losses (${\tan\delta<5\cdot10^{-8}}$ at 9~GHz below 90~K) and relatively high permittivity (${\varepsilon_\parallel\simeq11.5}$, ${\varepsilon_\perp\simeq9.5}$) \cite{braginsky1987experimental} allows for negligible field density on the Cu walls. We note that the choice of Cu is dictated by the need to work in magnetic fields: superconducting cavities are ruled out. This constraint is detrimental to the sensitivity at low $R_s$ values: our setup does not reach the sensitivity needed to assess the residual $R_s$ at low temperature, but is instead suitable for the high--R$_s$ regime typical of the vortex motion.

The measurements are performed in helium flow by slowly raising the temperature (0.1~K/min) after Field Cooling (FC) to the lowest temperature (typically 6~K). The field $H$ is applied perpendicular to the flat face of the sample.

Finally, in our experimental setup, the peak RF magnetic field amplitude parallel to the surface of the sample is assessed  to be $<20\;\mu$T. The low RF field amplitude allows a characterization of the surface impedance in the linear regime where the \Zs\ does not depend on the power of the applied RF field \cite{weinstock2012microwave}. It should be noted that the surface impedance in superconductors increases with the RF field amplitude \cite{martinello2016effect} and in \NbSn\ this trend is particularly accentuated (e.g.  in \NbSn\ the vortex dissipation due to the trapped field increases faster than what is observed in clean Nb)  \cite{Hall:IPAC2017-MOOCA2,martinello2016effect}.

\section{Results and discussion}
\label{results}

The flat polycrystalline bulk \NbSn~sample, of approximate dimensions  7~mm~$\times$ 5~mm, and 1~mm thick,  was obtained by sintering Nb and Sn powder (25~at.\%Sn) mixture under an Argon pressure of 2~kbar at 1250~\Celsius~in Hot Isostatic Pressure (HIP) conditions. Through X-ray diffraction methods (Rietveld refinement), the long-range order parameter $S$ was measured, showing a state of atomic ordering close to perfect ordering (${S=0.98\pm0.02}$) \cite{Spinathesis}.

\begin{figure}[!t]
\centering
\includegraphics[width=1\columnwidth]{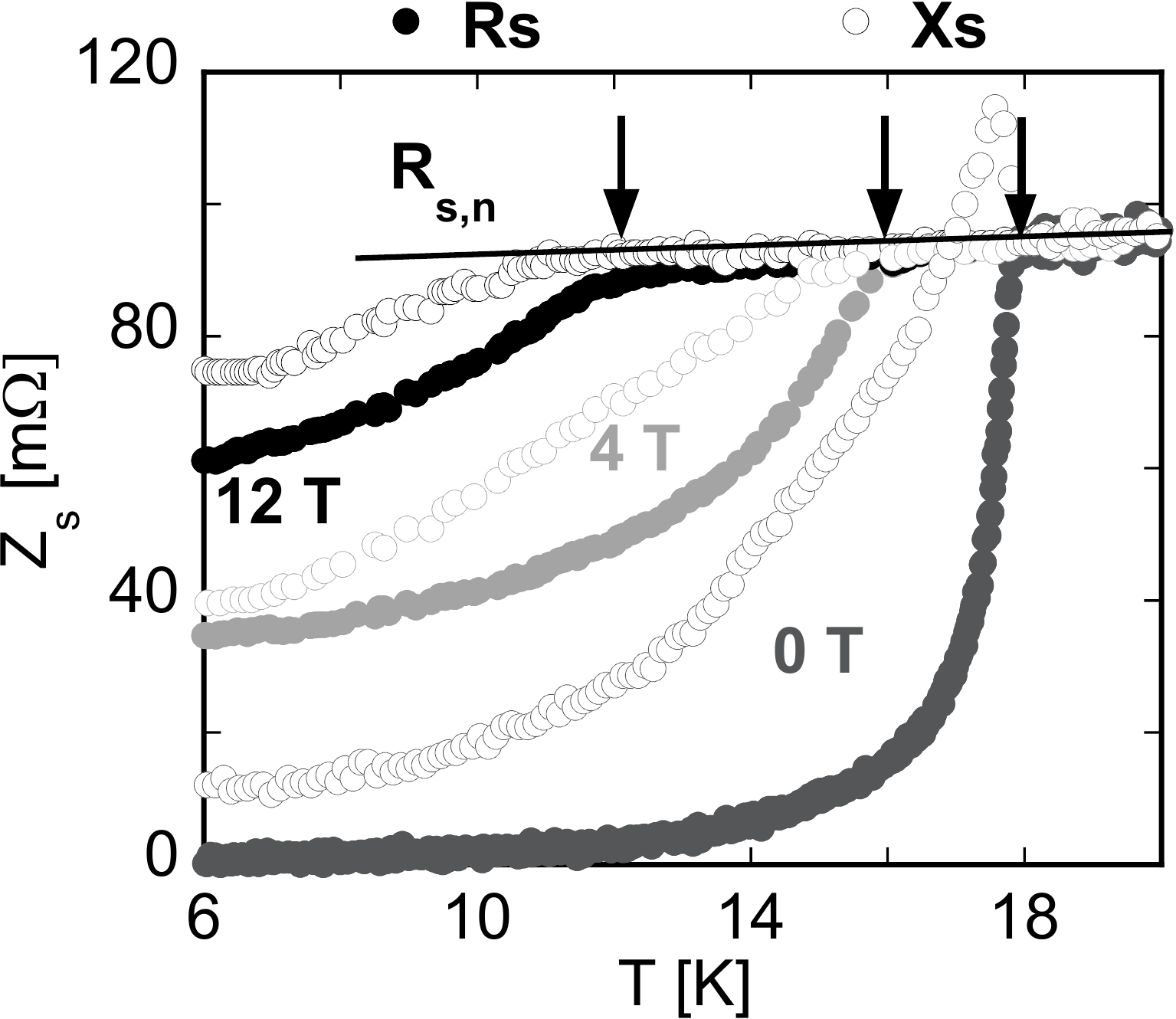}
\caption{$Z_s$ measured at fixed ${\mu_0H}$, depicted in the figure, and varying $T$. Full circles: $R_s$; empty circles: $X_s$. Colors:  black, ${\mu_0H=12}$~T; light gray, ${\mu_0H=4}$~T; dark gray, ${\mu_0H=0}$~T.  Vertical arrows indicate ${T_{c2}(H)}$.}
\label{fig:zs}
\end{figure}

In order to check the consistency of the data on our samples with the literature, we derived the normal state resistivity from
the \NbSn~surface resistance $R_{s,n}$ measured above $T_c$ (see \figurename \ref{fig:zs}): ${\rho_n=2R_{s,n}^2/\omega\mu_0=14.8\;\mu\Omega}$cm. The measured $\rho_n$ is typical of \NbSn~samples with $S=0.97$ and 25~at.\%Sn \cite{flukiger1987effect, godeke2006review}, entirely consistent with our results \cite{Spinathesis}. The penetration depth ${\lambda(0)=X_s(H=0,T\rightarrow0)/\mu_0\omega\sim100}$~nm is evaluated by extrapolating ${X_s(\mu_0H=0}$~T) at low temperature. The obtained value is in fair agreement with reported data \cite{li2017glag}.

The \NbSn~surface impedance, measured in FC at ${\mu_0H=\{0,\;4,\;12\}}$ T, is shown in \figurename \ref{fig:zs}. The beginning of the resistive transitions as a function of $T$ and $H$ are highlighted by the vertical arrows. We found that the behavior of the so-obtained ${H_{c2}(T)}$ is perfectly linear, and we estimate the derivative of the upper critical field $H_{c2}$ near $T_c$, ${\mu_0dH_{c2}/dT=2.03}$~T/K. The obtained value is fully consistent with literature \cite{PhysRevB.19.4545}.

\begin{figure}[!t]
\centering
\includegraphics[width=1\columnwidth]{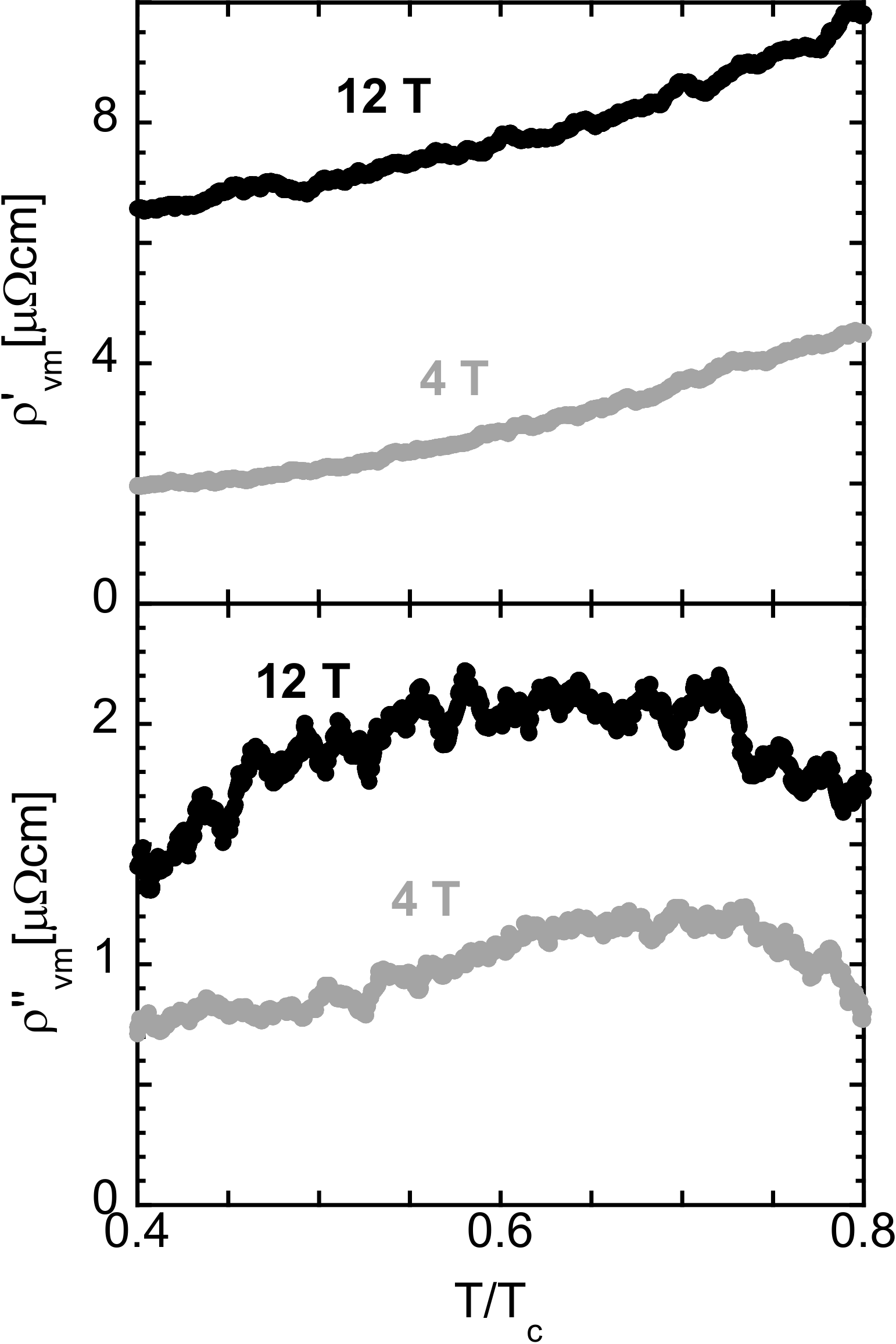}
\caption{The vortex motion resistivity ${\rho_{vm}=\rho'_{vm}+i\rho''_{vm}}$ obtained by the ${\Delta Z_s}$. In the upper plot ${\rho'_{vm}}$ and in the lower ${\rho''_{vm}}$. The black curves are measured at ${\mu_0H=12}$~T while the gray at ${\mu_0H=4}$~T. }
\label{fig:rhovm}
\end{figure}

\figurename~\ref{fig:zs} reports the set of measurements of ${Z_s(T,H)}$. The data do not present anomalous features. It can be deduced already from the raw data that the depinning frequency is of the same order of magnitude of the measuring frequency. In fact, the \textit{increases} of $R_s$ and $X_s$ with the field are different, although not much. Recalling \eqref{eqn:Zsrho}, \eqref{eq:GR}, one see that for both ${\nu\ll \nu_p}$ and ${\nu \gg \nu_p}$. the increase of  $R_s$ and $Z_s$ should be the same. We then focus on the variations ${\Delta Z_s= Z_{s}-Z_s(H=0)}$ in the temperature range ${T/T_c\lesssim 0.8}$, in order to avoid the high--$T$ region, where thermal effects introduce additional phenomena and then model parameters. We note in passing that working with the differences allows to alleviate the potential issues concerning the sensitivity of the resonant frequency to thermal expansion. We set, consistent with our results, ${\lambda(T)=\lambda_0/\sqrt{1-(T/T_c)^4}}$, and then from ${\Delta Z_s}$ we isolate ${\Re(\rho_{vm})=\rho_{vm}'}$ and ${\Im(\rho_{vm})=\rho_{vm}''}$. The vortex motion complex resistivity is reported in \figurename~\ref{fig:rhovm}. It is clearly seen that ${\rho_{vm}' > \rho_{vm}''}$, although the latter is non negligible. It can be also observed that ${\rho_{vm}''}$ shows a tendency to decrease at high $T$: this is completely reasonable, since at $T_c$ one has no imaginary part in the resistivity. Accordingly, ${\rho_{vm}'}$ steadily increases with $T$. We recall that, should it be pure flux--flow (no imaginary part), the well--known Bardeen--Stephen model \cite{PhysRev.140.A1197} predicts ${\rho_{vm}'\propto(T_c-T)^{-1}}$, consistent with our data.

\begin{figure}[!t]
\centering
\includegraphics[width=1\columnwidth]{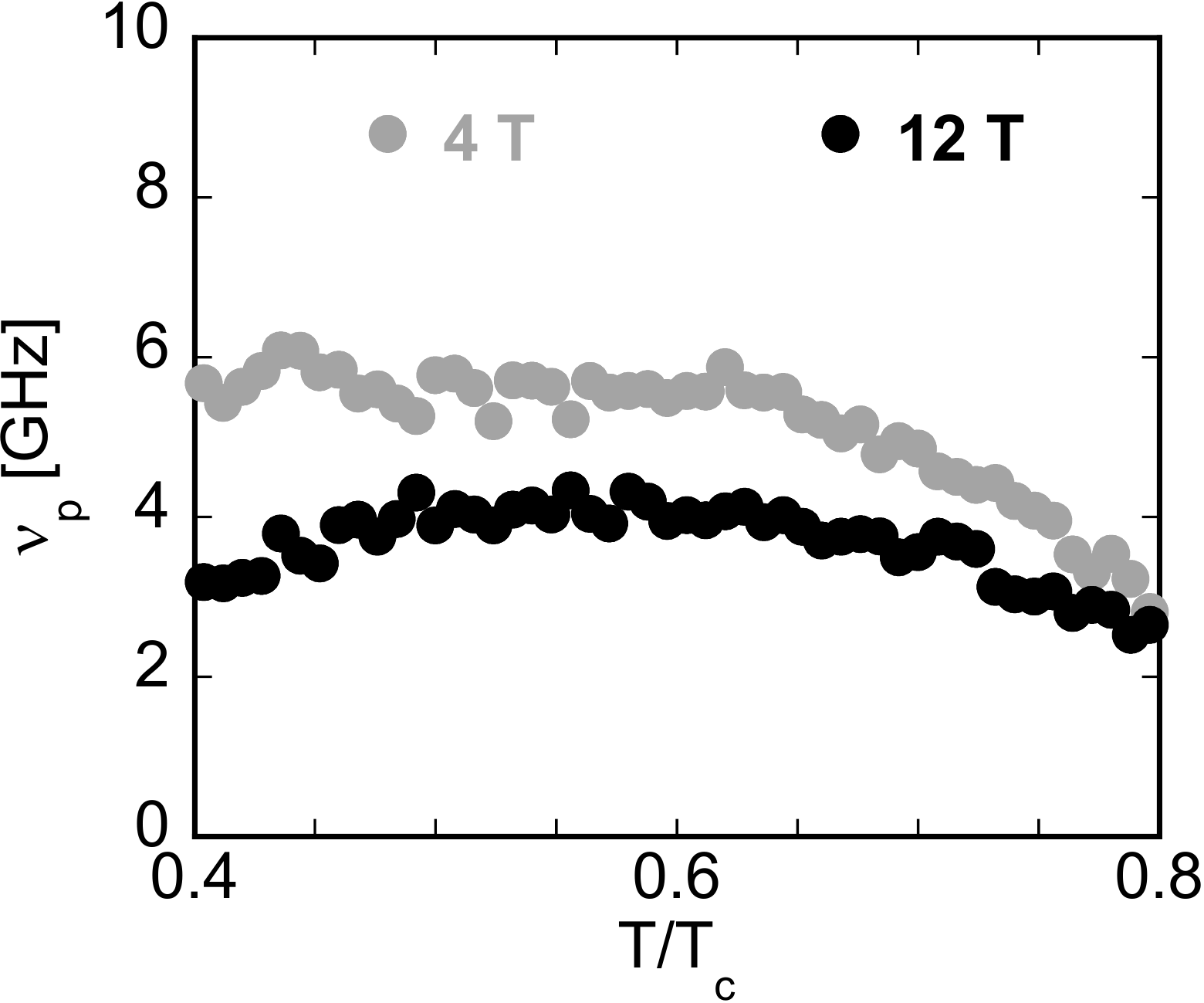}
\caption{The depinning frequency $\nu_p$ at ${\mu_0H=12}$~T (black points) and at ${\mu_0H=4}$~T (gray points).  The $\nu_p$ is almost constant up to 0.65~$T_c$. }
\label{fig:nu_p}
\end{figure}

From \eqref{eq:GR} we directly derive $\nu_p$ from the data in \figurename~\ref{fig:rhovm}. The data for $\nu_p$ are reported in \figurename~\ref{fig:nu_p}. We immediately note that the values for $\nu_p$ are quite large, ranging at low $T$ from $\sim$6~GHz at 4~T to 4~GHz at 12~T. These values compare very favourably to Nb. In pure Nb films $\nu_p$ rises with the decrease of film thickness, up to ${\nu_p\sim20}$~GHz in 10~nm film in 0.2~T perpendicular field, but it sharply falls down to 1~GHz in 160~nm films at 5~ K \cite{janjuvsevic2006microwave}. It can be deduced that $\nu_p$ in thick Nb films or bulks lays at best at 1~GHz, and more likely, well below. A second relevant aspect is the field resilience exhibited by \NbSn: $\nu_p$ is in the several GHz range in fields as high as 12~T, so that $\nu_p$ at 12~T in bulk \NbSn~$\nu_p$ is almost 5 times that of Nb thin film (thickness 160~nm) below 1~T, demonstrating enhanced \NbSn~RF behavior with respect to elementary Nb thin films \cite{janjuvsevic2006microwave, Pompeo2013,silva2011wideband}. Further improvements on \NbSn~$\nu_p$ are realistically reachable with \NbSn~thin films, where $\nu_p$ is expected to rise in analogy to Nb, opening interesting possibilities of \NbSn~applications at microwave frequencies. In order to complete the comparison with other superconductors, we note that a comparable depinning frequency (5~GHz) is exhibited by 13~$\mu$m thickness foils of Pb\textsubscript{0.83}In\textsubscript{0.17} at 1.7~K and 0.5~$H_{c2}$ \cite{gittleman1966radio}. Cuprates are known to have large depinning frequency \cite{golosovsky1996high,tsuchiya2001electronic}, with high values in YBa\textsubscript{2}Cu\textsubscript{3}O\textsubscript{7-x} (YBCO) single crystal, about ${\nu_p\sim20}$~GHz  at 45~K \cite{golosovsky1996high,tsuchiya2001electronic}. YBCO thin films with BaZrO\textsubscript{3} columnar and elongated defects exhibits still enhanced pinning frequencies: ${\nu_p\sim50}$~GHz at 70~K \cite{torokhtii2016measurement}. However, it must be mentioned that, at least in Tl$_2$Ba$_2$Ca$_2$CuO$_{8+x}$, such high $\nu_p$ are anomalously accompanied by a very large dissipation \cite{tallio2018}.

We make a final note on the possible effects of thermal activation (flux--creep). Should it be present, the effect is to reduce ${\rho_{vm}''}$ and to increase ${\rho_{vm}'}$: the more the thermal creep is prominent the more the pinning effect vanishes and ${\rho_{vm}\rightarrow\rho_{ff}}$ \cite{coffey1991unified,pompeo2008reliable}. From \eqref{eq:GR} and measurement frequency ${\nu>\nu_p}$, it can be shown that this phenomenon is modeled as a first approximation with a lowering of $\nu_p$ in \eqref{eq:GR} in order to obtain the same creep-induced increase in ${\rho_{vm}'}$ and reduction of ${\rho_{vm}''}$. Thus within this model, the obtained ${\nu_p}$  is at worst an underestimate. For this reason, we can state that our measurements set a lower limit to the \NbSn\ depinning frequency (\figurename~\ref{fig:nu_p}) without the need for entering any other parameter (i.e. the creep factor). Therefore, this result is robust against the possible presence of flux--creep: the depinning frequency remains rather high in \NbSn, much higher than in Nb.

\section{Conclusion}
\label{conc}
 
We presented  the first microwave (${\sim 15}$~GHz) characterization of \NbSn~in high magnetic fields (up to 12~T). The \NbSn~surface impedance \Zs~was evaluated in the mixed state to obtain information about the dissipative and reactive phenomena of vortex motion. From \Zs~elaborations we obtained and showed the vortex motion complex resistivity, a quantity that directly allowed us to obtain the depinning frequency $\nu_p$ of \NbSn. The depinning frequency is a parameter of great relevance because it establishes the frequency above which the elastic RF vortex motion becomes purely resistive. Hence, $\nu_p$ is the highest theoretical working frequency for low loss RF applications in high magnetic fields. 

The measured $\nu_p$ is almost constant in temperature (up to ${0.65\;T_c}$) and decreases from 6.0~GHz at 4~T, to 4.5~GHz at 12~T. This result shows that the \NbSn~exhibits much better RF characteristics than Nb and encourages the use of \NbSn~in technological RF and mw applications up to few GHz also in presence of dc magnetic fields.

\ifCLASSOPTIONcaptionsoff
  \newpage
\fi


\begin{thebibliography}{10}
\providecommand{\url}[1]{#1}
\csname url@samestyle\endcsname
\providecommand{\newblock}{\relax}
\providecommand{\bibinfo}[2]{#2}
\providecommand{\BIBentrySTDinterwordspacing}{\spaceskip=0pt\relax}
\providecommand{\BIBentryALTinterwordstretchfactor}{4}
\providecommand{\BIBentryALTinterwordspacing}{\spaceskip=\fontdimen2\font plus
\BIBentryALTinterwordstretchfactor\fontdimen3\font minus
  \fontdimen4\font\relax}
\providecommand{\BIBforeignlanguage}[2]{{%
\expandafter\ifx\csname l@#1\endcsname\relax
\typeout{** WARNING: IEEEtran.bst: No hyphenation pattern has been}%
\typeout{** loaded for the language `#1'. Using the pattern for}%
\typeout{** the default language instead.}%
\else
\language=\csname l@#1\endcsname
\fi
#2}}
\providecommand{\BIBdecl}{\relax}
\BIBdecl

\bibitem{xu2017review}
X.~Xu, ``{A review and prospects for Nb\textsubscript{3}Sn superconductor
  development},'' \emph{Supercond. Sci. Technol.}, vol.~30, no.~9, p. 093001,
  Aug 2017.

\bibitem{thome1994design}
R.~J. Thome, I.~J. Central, and H.~Teams, ``Design \& development of the {ITER}
  magnet system,'' \emph{Cryogenics}, vol.~34, pp. 39--46, May. 1994.

\bibitem{posen2017nb3sn}
S.~Posen and D.~L. Hall, ``{Nb\textsubscript{3}Sn} superconducting
  radiofrequency cavities: fabrication, results, properties, and prospects,''
  \emph{Supercond. Sci. Technol.}, vol.~30, no.~3, p. 033004, Jan. 2017.

\bibitem{valente2016superconducting}
A.~M. Valente-Feliciano, ``Superconducting {RF} materials other than bulk
  niobium: a review,'' \emph{Supercond. Sci. Technol.}, vol.~29, no.~11, p.
  113002, Sep. 2016.

\bibitem{padamsee1998rf}
H.~Padamsee, J.~Knobloch, and T.~Hays, \emph{{RF Superconductivity for
  Accelerators}}.\hskip 1em plus 0.5em minus 0.4em\relax Wiley-VCH, 1998.

\bibitem{flukiger2012overview}
R.~Fl{\"u}kiger, ``Overview of superconductivity and challenges in
  applications,'' \emph{Rev. Accel Sci. Technol.}, vol.~5, pp. 1--23, Apr.
  2012.

\bibitem{padamsee2009rf}
H.~Padamsee, \emph{{RF} superconductivity: science, technology, and
  applications}.\hskip 1em plus 0.5em minus 0.4em\relax John Wiley \& Sons,
  2009.

\bibitem{becker2015analysis}
C.~Becker, S.~Posen \emph{et~al.}, ``Analysis of {Nb\textsubscript{3}Sn}
  surface layers for superconducting radio frequency cavity applications,''
  \emph{Appl. Phys. Lett.}, vol. 106, no.~8, p. 082602, Feb. 2015.

\bibitem{grassellino2018accelerating}
A.~Grassellino, A.~Romanenko \emph{et~al.}, ``{Accelerating fields up to 49
  MV/m in TESLA-shape superconducting RF niobium cavities via 75C vacuum
  bake},'' \emph{arXiv preprint arXiv:1806.09824}, Jun. 2018.

\bibitem{gittleman1966radio}
J.~I. Gittleman and B.~Rosenblum, ``Radio-frequency resistance in the mixed
  state for subcritical currents,'' \emph{Phys. Rev. Lett.}, vol.~16, no.~17,
  p. 734, Apr. 1966.

\bibitem{collin2007foundations}
R.~E. Collin, \emph{Foundations for microwave engineering}.\hskip 1em plus
  0.5em minus 0.4em\relax John Wiley \& Sons, 2007.

\bibitem{coffey1991unified}
M.~W. Coffey and J.~R. Clem, ``Unified theory of effects of vortex pinning and
  flux creep upon the {RF surface impedance of type-II superconductors},''
  \emph{Phys. Rev. Lett.}, vol.~67, no.~3, p. 386, 1991.

\bibitem{tinkham1996introduction}
M.~Tinkham, \emph{Introduction to superconductivity}.\hskip 1em plus 0.5em
  minus 0.4em\relax Courier Corporation, 1996.

\bibitem{pompeo2008reliable}
N.~Pompeo and E.~Silva, ``Reliable determination of vortex parameters from
  measurements of the microwave complex resistivity,'' \emph{Phys. Rev. B},
  vol.~78, no.~9, p. 094503, Sep. 2008.

\bibitem{chen2004microwave}
L.-F. Chen, C.~Ong \emph{et~al.}, \emph{Microwave electronics: measurement and
  materials characterization}.\hskip 1em plus 0.5em minus 0.4em\relax John
  Wiley \& Sons, 2004.

\bibitem{hakki1960dielectric}
B.~Hakki and P.~Coleman, ``A dielectric resonator method of measuring inductive
  capacities in the millimeter range,'' \emph{IRE Trans. Microwave Theory and
  Tech.}, vol.~8, no.~4, pp. 402--410, Sep. 1960.

\bibitem{braginsky1987experimental}
V.~Braginsky, V.~Ilchenko, and K.~S. Bagdassarov, ``Experimental observation of
  fundamental microwave absorption in high-quality dielectric crystals,''
  \emph{Phys. Lett. A}, vol. 120, no.~6, pp. 300--305, Mar. 1987.

\bibitem{weinstock2012microwave}
H.~Weinstock and M.~Nisenoff, \emph{Microwave superconductivity}.\hskip 1em
  plus 0.5em minus 0.4em\relax Springer Science \& Business Media, 2012, vol.
  375.

\bibitem{martinello2016effect}
M.~Martinello, A.~Grassellino \emph{et~al.}, ``Effect of interstitial
  impurities on the field dependent microwave surface resistance of niobium,''
  \emph{Appl. Phys. Lett.}, vol. 109, no.~6, p. 062601, Aug. 2016.

\bibitem{Hall:IPAC2017-MOOCA2}
D.~Hall, J.~Kaufman \emph{et~al.}, ``{F}irst results from new single-cell
  {N}b\textsubscript{3}{S}n cavities coated at {C}ornell {U}niversity,'' in
  \emph{Proc. of International Particle Accelerator Conference (IPAC'17)}, May
  2017, pp. 40--43.

\bibitem{Spinathesis}
T.~Spina, \emph{Proton irradiation effects on Nb\textsubscript{3}Sn wires and
  thin platelets in view of High Luminosity LHC upgrade}.\hskip 1em plus 0.5em
  minus 0.4em\relax Ph.D thesis in Physics, Universite de Geneve, Deparement de
  Physique de la Matiere Quantique (DQMP), 2015.

\bibitem{flukiger1987effect}
R.~Flukiger, H.~Kupfer \emph{et~al.}, ``Effect of atomic ordering and
  composition changes on the electrical resistivity of{ Nb\textsubscript{3}Al,
  Nb\textsubscript{3}Sn, Nb\textsubscript{3}Ge, Nb\textsubscript{3}Ir,
  V\textsubscript{3}Si and V\textsubscript{3}Ga},'' \emph{IEEE Trans. Magn.},
  vol.~23, no.~2, pp. 980--983, Mar. 1987.

\bibitem{godeke2006review}
A.~Godeke, ``A review of the properties of {Nb\textsubscript{3}Sn and their
  variation with A15} composition, morphology and strain state,''
  \emph{Supercond. Sci. Technol.}, vol.~19, no.~8, p. R68, Jun. 2006.

\bibitem{li2017glag}
Y.~Li and Y.~Gao, ``{GLAG theory for superconducting property variations with
  A15 composition in Nb\textsubscript{3}Sn wires},'' \emph{Sci. Rep.}, vol.~7,
  no.~1, p. 1133, Apr. 2017.

\bibitem{PhysRevB.19.4545}
T.~P. Orlando, E.~J. McNiff \emph{et~al.}, ``{Critical fields, Pauli
  paramagnetic limiting, and material parameters of Nb\textsubscript{3}Sn and
  ${\mathrm{V}}_{3}$Si},'' \emph{Phys. Rev. B}, vol.~19, pp. 4545--4561, May.
  1979.

\bibitem{PhysRev.140.A1197}
J.~Bardeen and M.~J. Stephen, ``Theory of the motion of vortices in
  superconductors,'' \emph{Phys. Rev.}, vol. 140, pp. A1197--A1207, Nov. 1965.

\bibitem{janjuvsevic2006microwave}
D.~Janju{\v{s}}evi{\'c}, M.~S. Grbi{\'c} \emph{et~al.}, ``Microwave response of
  thin niobium films under perpendicular static magnetic fields,'' \emph{Phys.
  Rev. B}, vol.~74, no.~10, p. 104501, Sep. 2006.

\bibitem{Pompeo2013}
N.~Pompeo, K.~Torokhtii \emph{et~al.}, ``{Superconducting and Structural
  Properties of Nb/PdNi/Nb Trilayers},'' \emph{J. Supercond. Novel Magn.},
  vol.~26, no.~5, pp. 1939--1943, May. 2013.

\bibitem{silva2011wideband}
E.~Silva, N.~Pompeo, and S.~Sarti, ``Wideband microwave measurements in
  {Nb/Pd\textsubscript{84}Ni\textsubscript{16}/Nb structures and comparison
  with thin Nb films},'' \emph{Supercond. Sci. Technol.}, vol.~24, no.~2, p.
  024018, Jan. 2011.

\bibitem{golosovsky1996high}
M.~Golosovsky, M.~Tsindlekht, and D.~Davidov, ``High-frequency vortex dynamics
  in {YBa\textsubscript{2}Cu\textsubscript{3}O\textsubscript{7}},''
  \emph{Supercond. Sci. Technol.}, vol.~9, no.~1, p.~1, Sep. 1996.

\bibitem{tsuchiya2001electronic}
Y.~Tsuchiya, K.~Iwaya \emph{et~al.}, ``Electronic state of vortices in
  {YBa\textsubscript{2}Cu\textsubscript{3}O\textsubscript{y}} investigated by
  complex surface impedance measurements,'' \emph{Phys. Rev. B}, vol.~63,
  no.~18, p. 184517, Apr. 2001.

\bibitem{torokhtii2016measurement}
K.~Torokhtii, N.~Pompeo \emph{et~al.}, ``Measurement of vortex pinning in {YBCO
  and YBCO/BZO} coated conductors using a microwave technique,'' \emph{IEEE
  Trans. Appl. Supercond.}, vol.~26, no.~3, p. 8001605, Apr. 2016.

\bibitem{tallio2018}
N.~Pompeo, H.~Schneidewind, and E.~Silva, ``Measurements of microwave vortex
  response in dc magnetic fields in
  {Tl\textsubscript{2}Ba\textsubscript{2}CaCu\textsubscript{2}O\textsubscript{8+x}}
  films,'' presented at ASC 2018, abstract no. 2985707, submitted for
  publication.

\end{thebibliography}
\end{document}